\documentclass[a4paper,10pt]{article}
\usepackage[]{latexsym,amssymb,amsmath}
\newcommand{\be}{\begin{eqnarray}}
\newcommand{\ee}{\end{eqnarray}}
\newcommand{\bra}[1]{\mbox{$\langle\, #1 \mid$}}

\newcommand{\ket}[1]{\mbox{$\mid #1\,\rangle$}}

\newcommand{\pro}[2]{\mbox{$\langle\, #1 \mid #2\,\rangle$}}

\renewcommand{\d}{\mbox{${\rm d}$}}

\newcommand{\ve}{\mathbf}
\newcommand{\T}{\tau}
\title{\bf On the Unruh effect in de~Sitter space}
\author{R.~Casadio$^{ab}$\thanks{casadio@bo.infn.it},
S.~Chiodini$^a$\thanks{chiodo85@libero.it},
A.~Orlandi$^{ab}$\thanks{orlandi@bo.infn.it},
G.~Acquaviva$^{c}$\thanks{acquaviva@science.unitn.it}
$\ $,
R.~Di~Criscienzo$^{c}$\thanks{rdicris@science.unitn.it}
$\ $
and L.~Vanzo$^{c}$\thanks{vanzo@science.unitn.it}\\
\\
\\
$^a$Dipartimento di Fisica, Universit\`a di Bologna,
via~Irnerio~46, 40126~Bologna, Italy
\\
$^b$
I.N.F.N.,
Sezione di Bologna, via~Irnerio~46, 40126~Bologna, Italy
\\
$^c$Dipartimento di Fisica, Universit\`a di Trento,
\\
and I.N.F.N., Gruppo Collegato di Trento,
via Sommarive~14, 38100~Povo, Italy}
\begin{document}
\maketitle
\begin{abstract}
\noindent We give an interpretation of the temperature in de~Sitter universe in terms of a dynamical
Unruh effect associated with the Hubble sphere.
As with the quantum noise perceived by a uniformly accelerated observer in static
space-times, observers endowed with a proper motion can in principle detect the effect.
In particular, we study a ``Kodama observer'' as a two-field Unruh detector for
which we show the effect is approximately thermal. 
We also estimate the back-reaction of the emitted radiation and find trajectories
associated with the Kodama vector field are stable.

\end{abstract}
\section{Introduction}
\setcounter{equation}{0}
In Ref.~\cite{kodama}, Kodama introduced a vector field which generalizes
the notion of time-like Killing vectors to space-times with dynamical
horizons~\footnote{For a clarification of the differences between Kodama and
Killing vectors, see Ref.~\cite{KvK}.
For the ability of Kodama vectors to define  a local time flow, see Ref.~\cite{Abreu:2010ru}}.
Recently many authors have focused their attention on the Kodama vector field
and found interesting results related to the measurement of the Hawking radiation
emitted by different kinds of dynamical horizons~\cite{Abreu:2010ru,Hayward:1997jp,Hayward:1998ee,Racz:2005pm,
DiCriscienzo:2007fm, DiCriscienzo:2008dm,Hayward:2008jq,Nielsen:2005af,DiCriscienzo:2010zza,
Nielsen:2008kd,Wu:2008ir,chen} or time-like naked singularities~\cite{DiCriscienzo:2010gh}.
\par
Here we present a more ``physical'' process of measurement in which the detector
has finite mass and extension~\cite{casadio95}, and interacts with a massless scalar field
like in Ref.~\cite{casadio99}.
The background is a de~Sitter universe and the horizon is provided by its exponential expansion.
The Kodama observer does not follow a cosmic fluid geodesic, but remains
at constant proper distance from the horizon, and is thus ``accelerated''
with respect to locally inertial observers.
Consequently, it perceives the de~Sitter vacuum as a thermal bath with a temperature
associated with the surface gravity of the Hubble sphere, much like
a uniformly accelerated detector in Minkowski space-time measures a temperature
associated with its acceleration.
As usual, such result is obtained from the transition probability for the detector to emit a
(scalar) radiation quantum and increase its energy.
Further, assuming finite detector's size and mass, we show that the Kodama trajectory
remains stable during this process.
\par
We shall use the metric signature $(+ -  -  -)$ and implement  units where $c=G=k_B=1$.
\section{Kodama observer in de~Sitter space}
\setcounter{equation}{0}
We recall the de~Sitter metric for an observer comoving with the cosmic fluid at
constant $r$ reads
\be
\d s^2
=
\d t^2-e^{2\,H\,t}\left(\d r^2 + r^2\, \d \Omega \right)
\ .
\label{dS}
\ee
We are here interested in an observer at fixed areal position which moves along
the trajectory
\be
r\,e^{Ht}=K
\ ,
\label{Ktraj}
\ee
with $K$ constant.
This observer stays at fixed distance from its cosmological horizon and
its four-velocity is given by
\be
u^\mu=\frac{k^\mu}{\sqrt{1-H^2K^2}}
\ ,
\label{velocity}
\ee
where $k^\mu=\left(1, -H\,r, 0, 0\right)$ is the Kodama vector related to the surface gravity
of the de~Sitter horizon
\be
2\,\pi\,T_H=\left.K^{-1}\right|_{K=H^{-1}}=H
\ .
\ee
We also note the observer's four-acceleration is
\be
a^\mu
=\left(
\frac{H^3\,K^2}{1-H^2\,K^2}
,
-\frac{H^2\,K\,e^{-H\,t}}{1-H^2K^2}, 0,0
\right)
\ ,
\label{a_r}
\ee
with modulus
\be
A^{2}
\equiv a^\mu\, a_\mu
=
-\frac{H^4\,K^2}{1-H^2\,K^2}
\ .
\label{acc}
\ee
One can also introduce coordinates naturally associated with the Kodama detector,
namely $K$ from Eq.~\eqref{Ktraj} and $T$, so that the metric reads
\be
\d s^2=\left(1-H^2\,K^2\right)\d T^2
+2\,H\,K\,\d T\,\d K
-\d K^2
-K^2\,\d\Omega^2
\ ,
\label{kodamametric}
\ee
which shows that the Kodama vector $k=\partial/\partial T$ is a Killing vector.
%
%
%
%
%
%
\section{Particle detectors}
\setcounter{equation}{0}
\label{detector}
The concept of particle becomes rather involved once gravity and curved space-times are taken
into account~\cite{Birrell:1982ix}.
In the following, we shall therefore employ the particle detector proposed by Unruh~\cite{unruh} and
DeWitt~\cite{dewitt}, which can be described as a quantum mechanical particle with
energy levels linearly coupled to a massless scalar field via a monopole moment operator.
\subsection{Point-like detectors}
\label{pointK1}
Let us consider a massless scalar field $\hat\phi$ with Hamiltonian $\hat H_\phi$ obeying
the massless Klein-Gordon equation with dispersion relation
$\omega_{\ve k} = \vert \ve k \vert$.
The \textit{point-like\/} detector is a quantum mechanical system with a set of energy eigenstates
$\{\ket {E_i}\}$ which moves along a prescribed classical trajectory
$t= t(\T), \;\ve x= \ve x(\T)$, where $\T$ is the detector's proper time.
The detector is coupled to the scalar field $\hat\phi$ via the interaction Hamiltonian
\be
\hat H_{\rm int} = \lambda \,\hat M(\T) \hat\phi(\T) \label{int ham}
\ ,
\ee
with $\lambda$ a small parameter and $\hat M(\T)$ the detector's
monopole moment operator.
\par
Suppose that at time $\T_0$ the detector and field are in the product state
$\ket {0, E_0} = \ket 0 \ket {E_0}$, where $\ket{0}$ is the scalar field vacuum and
$\ket {E_0}$ a detector state with energy $E_0$~\cite{Langlois:2005if}.
We want the probability that at a later time $\T_1 > \T_0$ the detector
is found in the state $\ket {E_1}$ with energy $E_1 \gtrless E_0$,
regardless of the final state of the field $\hat\phi$.
The answer can be readily obtained in the so-called interaction picture where
states are evolved by the Schr\"{o}dinger equation with the interaction Hamiltonian
$\hat H_{\rm int}$.
To first order in perturbation theory,
the amplitude for the transition from $\ket {0, E_0}$ at time $\T=\T_0$
to $\ket {\varphi, E_1}$ at time $\T=\T_1$ is then
\be
\pro{\varphi, E_1}{0, E_0}
&\!\!=\!\!&
\bra{\varphi, E_1} \hat T \exp \left(- i \int_{\T_0}^{\T_1} d\T \, \hat H_{\rm int} (\T)\right) \ket{0, E_0}
\nonumber
\\
&\!\!\simeq\!\!&
- i\, \lambda \,\bra{E_1}\hat M(0)\ket{E_0}
\int_{\T_0}^{\T_1} d\T\, e^{i\, \T\, (E_1 - E_0)} \bra{\varphi} \hat \phi(\T)\ket{0}
\ .
\label{amplit}
\ee
where $\hat T$ is the time-ordering operator.
Squaring~(\ref{amplit}) and summing over all final field states, yields
\be
\sum_\varphi \vert \pro{\varphi, E_1}{0, E_0} \vert^2
=
\lambda^2  \vert \bra{E_1}\hat M(0)\ket{E_0} \vert^2
\int_{\T_0}^{\T_1} d\T
\int_{\T_0}^{\T_1} d\T'\, e^{- i (E_1 -E_0) (\T - \T')}
\bra{0} \hat \phi(\T)\, \hat \phi(\T') \ket{0}
\label{trans}
\ . 
\ee
The pre-factor $\lambda^2  \vert \bra{E_1}\hat M(0)\ket{E_0} \vert^2$
depends only on the peculiar details of the detector, whereas the
\textit{response function} 
\be
R_{\T_0, \T_1} (\Delta E)
=
\int_{\T_0}^{\T_1} d\T \int_{\T_0}^{\T_1} d\T'\,
e^{- i \Delta E (\T - \T')} \bra{0} \hat \phi(\T)\, \hat \phi(\T') \ket{0}
\label{resp}
\ee
is the same for all possible detectors.
Here, we have set the energy gap $\Delta E \equiv E_1 - E_0 \gtrless 0$
for excitations or decay, respectively.
>From now on, we will only consider the model-independent response function.
\par
Introducing new coordinates $u:= \T,\, s:= \T -\T'$ for $\T > \T'$ and
$u:= \T',\, s:= \T'-\T$ for $\T'>\T$, the response function can be re-written as
\be
R_{\T_0, \T_1} (\Delta E)
=
2 \int_{\T_0}^{\T_1} du\,\int_{0}^{u-\T_0} d s
\; \mbox{Re} \left(e^{- i \Delta E s} \bra{0} \hat \phi(u)\, \hat \phi(u-s) \ket{0}\right)
\ ,
\label{resp2}
\ee
having used $\bra{0} \hat \phi(\T') \hat \phi(\T) \ket{0} = \bra{0} \hat \phi(\T) \hat \phi(\T') \ket{0} ^*$,
since $\hat \phi$ is a self-adjoint operator.
Eq.~(\ref{resp2}) can be differentiated with respect to $\T_1$ in order to obtain
the \textit{transition rate}
\be
\dot R_{\T_0, \T} (\Delta E)
=
2 \int_{0}^{\T-\T_0} d s\;
\mbox{Re} \left(e^{- i \Delta E s} \bra{0} \hat \phi(\T)\, \hat \phi(\T-s) \ket{0}\right)
\ ,
\label{trans rate}
\ee 
where we set $\T_1 \equiv \T$.
If the correlation function $\bra{0} \hat \phi(\T) \hat \phi(\T-s) \ket{0}$
is invariant under $\T$-translations, Eq.~\eqref{trans rate} can be finally simplified
to
\be
\dot R_{\T_0, \T} (\Delta E)
=
\lim_{\epsilon \rightarrow 0^+}
\; \int_{-(\T-\T_0)}^{\T-\T_0} d s\; e^{- i \Delta E s} \,W_{\epsilon} (x(s),x(0))
\ .
\label{trans rate3}
\ee
where
\be
W_\epsilon(x(\T), x(\T'))
\equiv
\bra{0} \hat \phi(x(\T))\, \hat \phi(x(\T')) \ket{0}
=
\frac{1/4\pi^2}{\vert \ve x(\T) - \ve x(\T')\vert^2-[t(\T)-t(\T') - i \epsilon]^2}
\label{wightman}
\ee
is the positive frequency Wightman function.
\subsection{Unruh effect in de~Sitter universe}
\label{pointK2}
Let us apply the above construction to Kodama detectors in de~Sitter space-time.
Since Kodama observers are just the stationary de~Sitter observers
at constant distance from their cosmological horizon,
we know that they will perceive a thermal bath at de~Sitter temperature \cite{Gibbons:1977mu}
\be
T_H= \frac{H}{2\,\pi}
\ .
\label{TH}
\ee
\par
We can confirm this expectation by noting that de~Sitter space is conformally
flat with conformal time $\eta = -H^{-1} e^{-H\,t}$.
Then, provided $1-K^2H^2 > 0$,  the Wightman function~(\ref{wightman})
becomes
\be
W_\epsilon(x,x')
=
-\frac{1 }{4 \pi^2} \frac{H^2}{e^{H( t + t')}}\frac{1}{(1-K^2H^2)(e^{-Ht} - e^{-Ht'} - i\epsilon)^2 }
\ .
\ee
and, in the limit of $\T_0 \rightarrow -\infty$, the detector transition rate~(\ref{trans rate3})
reads
\be
\dot R(\Delta E)
=
-\frac{H^2}{4\pi^2} \lim_{\epsilon \rightarrow 0^+}
\;\int_{-\infty}^{+\infty} ds \; 
\frac{\exp( -i \Delta E \,\sqrt{1-K^2H^2}\, s)}{(e^{Hs/2} - e^{-Hs/2} - i\epsilon)^2}
\ ,
\ee
where we also used
\be
\tau = t\,\sqrt{1-K^2\,H^2}
\ .
\label{tau}
\ee
As long as $H>0$ (expanding universe) we can write the denominator as
\begin{equation}
\left(e^{H\,s/2} - e^{-H\,s/2} - i\epsilon\right)^2
=
4\,\sinh^2\left[ H\,(s-i\epsilon)/2\right]
\ .
\end{equation}
This function has infinitely many double poles in the complex $s$-plane,
namely for $s=s_j$ with
\begin{equation}
s_j = \frac{2\,\pi\, i}{H} j
\ ,
\ \ \
j \in \mathbb Z
\ .
\end{equation}
Since we are interested in the case of $\Delta E>0$, we can close the contour
of integration in the lower half plane, summing over the residues of all the double
poles in the lower complex $s$-plane, with the exception of the $s=0$ pole which
has been slightly displaced above the integration path by the $i\epsilon$-prescription.
The well known result turns out to be confirmed, namely
\be
\dot R(\Delta E)
= \frac{1}{2\pi} \cdot\frac{\Delta E\, \sqrt{1-K^2H^2}}{\exp\left( \frac{2\,\pi \,\Delta E\, \sqrt{1-K^2\,H^2}}{H} \right) - 1}
\ ,
\label{Rdesitter}
\ee
showing the presence of a cosmological Unruh effect with de~Sitter temperature~\eqref{TH}
red-shifted by the Tolman factor $\sqrt{1-K^2\,H^2}$, here appearing as a Doppler shift
due to the proper motion of the detector.
In fact, we remarked before that $K$ is also the static coordinate
of the detector relative to the static patch.
\par
Another intriguing formula can be written if we recall the acceleration~\eqref{acc}, namely
\be
\frac{T_H}{\sqrt{1-H^{2}\,K^{2}}}=\frac{\sqrt{A^{2}+H^{2}}}{2\pi}
\ .
\ee
The observed temperature here appears due to a mixing of
a pure Unruh effect (the acceleration term) plus a cosmological expansion term (the $H$ term),
and is the de~Sitter version~\cite{Narnhofer:1996zk} of a formula discovered by Deser and Levine
for detectors in anti-de~Sitter space~\cite{Deser:1998xb}.
\par
Alternatively, we can understand this effect as the transition from cosmological
energy $\Delta E$, conjugated to cosmic de~Sitter time $t$, to the energy
$\Delta E\,\sqrt{1-K^2\,H^2}$ as measured locally by Kodama's observers, with proper time
$\tau$ of Eq.~\eqref{tau}.
\subsection{Extended detector and back-reaction}
We now repeat the previous analysis assuming the detector's size is not
{\em a priori\/} negligible.
In order to simplify the computation, we still assume spherical symmetry
and the detector therefore only extends along one dimension.
We will not display all the details but focus on the main differences.
\par
It is easy to see that a detector moving along a Kodama trajectory~\eqref{Ktraj}
in an expanding de~Sitter universe,
has the same dynamics of a particle which moves along the separatrix in the potential of
an inverted harmonic oscillator.
>From the equation of motion~\eqref{Ktraj}, we can therefore introduce the effective Lagrangian
\be
L_{\rm iho}=m\left(\frac{\d^2 r}{\d t^2} + H^2\,r^2\right)
\ ,
\label{effL}
\ee
where $m$ is a parameter with mass dimension whose relation with physical quantities will 
be clarified later.
Our detector is initially (at $t=0$) represented by a Gaussian wave packet of
size $b$ peaked around $r=K$,
\be
\psi(0,r)=\frac{\exp\left[-i\strut\displaystyle\frac{H\,K\,m\,r}{\hbar}-\frac{(r-K)^2}{2\,b^2}\right]}
{\sqrt{b\,\sqrt{\pi}}}
\ ,
\ee
which is then propagated to later times by the propagator obtained from the
Lagrangian~\eqref{effL} (see, e.g., Ref.~\cite{casadio95}),
\be
G(t,r;0,r')=
\sqrt{\frac{i\,H\,m}{2\,\pi\,\hbar\,\sinh(H\,t)}}\,
\exp\left[
i\,\frac{H\,m\left[(r^2-r'^2)\,\cosh(H\,t)-2\,r\,r'\right]}{2\,\hbar\,\sinh(H\,t)}
\right]
\ .
\ee
The complete expression of the detector's propagated wavefunction
is rather cumbersome, however we notice that its square modulus
reads
\be
|\psi(t,r)|^2
\simeq
\exp\left[-\frac{2\,b^2\,H^2\,m^2\,(r-K\,e^{-H\,t})^2}
{b^4\,H^2\,m^2-\hbar^2+(\hbar^2+b^4\,H^2\,m^2)\,\cosh(H\,t)}\right]
\ ,
\label{psi2}
\ee
which represents a Gaussian wave-packet peaked along the classical trajectory~\eqref{Ktraj}
employed in Sections~\ref{pointK1}-\ref{pointK2} and spreading in time. 
The classical behavior is properly recovered in the limit $\hbar\to 0$ followed by
$b\to 0$~\cite{casadio95}, in which the detector's wavefunction $\psi(r,t)$ reproduces
the usual Dirac $\delta$-function peaked on the classical trajectory.
\par
Now, in order to study the probability for the detector to absorb a scalar quantum
and make a transition between two different trajectories (parameterized by different
$m_i$ and $K_i$), one needs to compute the transition amplitude for finite
$b$ and $\hbar$ (otherwise the result would automatically vanish, the response
function involving the product of two Dirac $\delta$'s peaked on different trajectories). 
The detector also interacts with the quantized scalar field $\varphi=\varphi(t,r)$ according to
\be
\mathcal{L}_{\rm int}=\frac{1}{2}\,Q\left(\psi_2^*\,\psi_1+ \psi_2\, \psi_1^* \right)\varphi
\ ,
\ee
where $Q$ is a coupling constant and $\psi_i=\psi_i(t, r)$ two possible states
of the detector corresponding to different trajectories $r_i=K_i\,e^{-H\,t}$
and mass parameters $m_i$~\footnote{A fundamental
difference with respect to the Unruh effect analyzed in Ref.~\cite{casadio99} is that
the acceleration parameter $H$ is not varied here, since it is a property of the background
space-time.
A change $\delta K$ implies a change in the detector's acceleration according to
Eq.~\eqref{acc}.}.
We assume the difference between the two states is small,
\be
\left\{
\begin{array}{l}
K_1=K-\frac{1}{2}\,\delta K
\\
\\
K_2=K+\frac{1}{2}\,\delta K
\end{array}
\right.
\quad
,
\qquad
\left\{
\begin{array}{l}
m_1=m-\frac{1}{2}\,\delta m
\\
\\
m_2=m+\frac{1}{2}\,\delta m
\ ,
\end{array}
\right.
\ee
and expand to leading order in $\delta K$ and $\delta m$ and,
subsequently, for short times ($H\,t\sim H\,t'\ll 1$), keeping $\hbar$ and $b$ finite.
In particular, one obtains
\be
\psi_2^*\,\psi_1(t)\,\psi_1^*\,\psi_2(t')
\simeq
\exp\left[-i\,\frac{H^2\,K^2}{\hbar}\,\delta m\,(t-t') + O(b) \right]
\ ,
\ee
in which we have evaluated the phase (in the saddle-point approximation) with $r$
along the average trajectory between $r_1$ and $r_2$~\cite{casadio99}.
Like the Unruh detector in Ref.~\cite{casadio99}, the above exponential does not contain
a real (quadratic in $\delta K$) part,
contrary to the case of a geodetic observer [see Eq.~\eqref{phaseGeo}
in Appendix~\ref{flat})], which implies that the transition amplitude will not
vanish in the point-like limit $b\to 0$.
Upon comparing with the result obtained for the point-like case,
we immediately recognize that
\be
H^2\,K^2\,m=E\,\sqrt{1-H^2\,K^2}
\ ,
\ee
where $E$ is the detector's proper energy and
\be
\psi_2^*\,\psi_1(t)\,\psi_1^*\,\psi_2(t')
\simeq
\exp\left[-\frac{i}{\hbar}\,\delta E\,\sqrt{1-H^2\,K^2}\,(t-t') 
+ \frac{i}{\hbar}\,\frac{2- 3 H^2\,K^2}{K \sqrt{1-K^2\,H^2}}\,E\,\delta K\,(t-t') + \mathcal{O}(b) \right]
\ .
\label{phaseK}
\ee
We can now take the limit $b\to 0$, as part of the point-like limit in which
one would not expect the second term in the above exponential. 
In Ref.~\cite{casadio99}, we required the analogue of the second term
above vanished and obtained the equation of motion for a uniformly accelerated
detector in Minkowski space-time, namely $m\,a\,=f$ and constant.
Following the same line of reasoning, we now obtain the equation of motion
\be
\delta K=0
\ .
\label{dK0}
\ee
This can be interpreted as meaning the Kodama trajectory is stable against
thermal emission of scalar quanta in the de~Sitter background.
\par
The transition probability per unit de~Sitter time can finally be computed
by taking the classical limit, in which one recovers the same result 
in Eq.~\eqref{Rdesitter} with $\Delta E=\delta E$.
\section{Conclusions}
\setcounter{equation}{0}
We considered an expanding de Sitter universe that is inflating exponentially and
studied the dynamics of an observer as an object which is collapsing towards the
Hubble sphere.
This is the ``Kodama observer'' placed at $K=r\, e^{H\,t}$ constant, and its
radial velocity~\eqref{velocity} is in fact negative which means that the observer is
moving towards decreasing radii.
In particular, the observer's motion is described by an inverted harmonic potential
and involves the negative radial acceleration in Eq.~(\ref{a_r}).
We can picture our observer/detector in spherical coordinates as a ``shell''
with a ``density'' profile $|\psi|^2$ peaked on the average radius $r$ [see Eq.~(\ref{psi2})]
which sees the universe becoming less dense (in time) but in a homogenous way (in space).
Of course, the detector's proper time $\tau$ in Eq.~\eqref{tau} differs from the cosmological time
$t$ for an observer comoving with the cosmic fluid. 
\par
The detector interacts with a scalar field so that it can absorb or emit quanta and change
its proper mass (energy).
Indeed, the Unruh effect is given by the simultaneous emission of a scalar quantum and
detector's excitation.
We therefore considered the transition between two states of the detector
corresponding to two different trajectories associated with different energies and
values of $K$ and obtained the transition rate \eqref{Rdesitter} in the point-like
limit.
Our result shows the expected thermal behavior~\eqref{TH} associated with the trajectory's stability
\eqref{dK0}.
\par
In summary, we studied the response of a Kodama detector moving along trajectories
at constant distance from the Hubble sphere of the de~Sitter universe and
found that it perceives a thermal noise associated with the emission
of scalar quanta.
We also estimated the back-reaction of the emitted radiation and showed
trajectories associated with the Kodama vector fields are stable.
This represents a novel semiclassical property of the (classically defined)
Kodama vector.
\appendix
\section{Extended detectors in flat space-time}
\label{flat}
In order to clarify the analysis of the extended Kodama detector in de~Sitter
space, let us review the case of a detector moving along a geodesic in
flat space-time.
Without loss of generality, we consider a detector at fixed $x=0$ in a Minkowskian
coordinate system $\{t,x,y,z\}$ which, for simplicity, extends only along the 
axis $x$. 
This case can also be interpreted as a detector comoving with the cosmic
fluid in a general Friedmann-Robertson-Walker metric by replacing $x$
with the radial comoving coordinate $r$.
\par
The initial wavefunction for such a detector is now 
\be
\psi(0,x)=\frac{\exp\left[-\frac{x^2}{2\,b^2}\right]}
{\sqrt{b\,\sqrt{\pi}}}
\ ,
\ee
and is propagated in time by the free particle propagator
\be
G(t,x;0,x')=
\sqrt{\frac{m}{2\,\pi\,i\,\hbar\,(t-t')}}\,
\exp\left[
i\,\frac{m\left(x^2-x'^2\right)}{2\,\hbar\,(t-t')}
\right]
\ ,
\ee
which yields
\be
\psi(t,x)=
\sqrt{\frac{b\,m}{i\,\sqrt{\pi}\left(\hbar\,t-i\,b^2\,m\right)}}
\exp\left[-\frac{m\,x^2}{2\left(b^2\,m+i\,\hbar\,t\right)}\right]
\ .
\ee
As before, we now consider the transition between two detectors
with slightly different positions and mass,
\be
\left\{
\begin{array}{l}
x_1=-\frac{1}{2}\,\delta x
\\
\\
x_2=\frac{1}{2}\,\delta x
\end{array}
\right.
\quad
,
\qquad
\left\{
\begin{array}{l}
m_1=m-\frac{1}{2}\,\delta m
\\
\\
m_2=m+\frac{1}{2}\,\delta m
\ ,
\end{array}
\right.
\ee
and expand to leading order in $\delta x$ and $\delta m$,
keeping $\hbar$ and $b$ finite.
In particular, by expanding in $\hbar$ for $b$ fixed, one obtains
\be
\psi_2^*\,\psi_1(t)\,\psi_1^*\,\psi_2(t')
\simeq
\exp\left[ -\frac{\delta x^2}{4\,b^2}
+i\,\frac{\hbar\,t\,\delta x^2\,\delta m}{8\,b^4\,m^2}+\mathcal{O}(\hbar^2)\right]
\ ,
\label{phaseGeo}
\ee
in which we have evaluated the phase for $x$ along the average trajectory
between $x_1$ and $x_2$ (that is, $x=0$).
Note that the first term in the exponential generates a Dirac-$\delta(\delta x)$ which
therefore forces the transition to occur at $\delta x=0$ as one would expect for
a geodetic motion with no emission or absorption.
In fact, the condition $\delta m=0$ is ensured by the Wightman's function evaluated
along geodesics.
Further, no terms proportional to $1/\hbar$ appear at all and, in particular,
none analogous to the second one in Eq.~\eqref{phaseK}.
%

%
%
%
%
%
\end{document}